# Vernacularizing Taxonomies of Harm is Essential for Operationalizing Holistic AI Safety

Wm. Matthew Kennedy[1] and Daniel Vargas Campos[2]

[1] University of Sussex
[2] Independent Researcher
vernacularizing.taxonomies@gmail.com

**Abstract**

Operationalizing AI ethics and safety principles and frameworks is essential to realizing the potential benefits and mitigating potential harms caused by AI systems. To that end, actors across industry, academia, and regulatory bodies have created formal taxonomies of harm to support operationalization efforts. These include novel "holistic" methods that go beyond exclusive reliance on technical benchmarking. However, our paper argues that such taxonomies must also be transferred into local categories to be readily implemented in sector-specific AI safety operationalization efforts, and especially in underresourced or "high-risk" sectors. This is because many sectors are constituted by discourses, norms, and values that "refract" or even directly conflict with those operating in society more broadly. Drawing from emerging anthropological theories of human rights, we propose that the process of "vernacularization"—a participatory, decolonial practice distinct from doctrinary "translation" (the dominant mode of AI safety operationalization)—can help bridge this gap. To demonstrate this point, we consider the education sector, and identify precisely how vernacularizing a leading holistic taxonomy of harm leads to a clearer view of how harms AI systems may cause are substantially intensified when deployed in educational spaces. We conclude by discussing the generalizability of vernacularization as a useful AI safety methodology.

## Introduction

AI systems are sociotechnical systems. They encode values and preferences found in human societies (Weinstein et al 2021, Barocas 2023). That means the harms or benefits they may create are socially and culturally constructed. They are not inherent to the technical aspects of the system. However, the vast majority of sociotechnical safety evaluations, metrics, and measurement methodologies examine only those technical aspects (Weidinger et al 2023). The result is a tremendous, perhaps crushing, accumulation of "ethical debt" that many AI ethics, safety, and governance teams are racing to pay down (Petrozzino 2021).

A promising approach to solving this pressing problem comes from value-sensitive design, responsible innovation, and trust and safety methodologies. From these fields have come several proposals for expanding governance functions and cultures–for instance by infusing technical product teams with ethics and safety personnel, including ethics and safety training curriculum in degree programs or during onboarding, or creating accountable teams dedicated to ethical review functions (Lange 2023). Moreover, researchers across AI labs and public entities alike have made significant efforts to advance our measurement science of harms caused by AI systems through proposing taxonomies of harms against which targeted technical safety evaluations can be conducted (Weidinger et al 2021; Bommanisi et al 2022, Weidinger et al 2022; Shelby et al 2023; Magooda et al 2023).

Recently, conceptual work arguing for *holistic* safety evaluations–that is, evaluations of harms arising at the technical capability, human-interaction, and broad-system levels–has made a compelling case for the urgent need to move beyond an exclusive reliance on technical benchmarking (Weidinger et al 2023). Such arguments are made more convincing considering that the vast majority of technical benchmarks are intended for LLMs, and do not account for multimodal models. Thanks to this work, it is now clearer than ever that *context* (sociocultural, interaction, and longitudinal) is essential to evaluating risks of harm that AI systems may cause.

Despite this clarity, operationalizing these frameworks and approaches has often proved exceedingly difficult. There are many reasons for this, including structural obstacles, internal resistance to or deprioritization of ethics and safety work, and uncertainty about the business value of implementing them (Gutierrez et al 2022, Bevilacqua et al 2023). Furthermore, operationalizing each of these general taxonomies has been regarded as the responsibility of sector-specific AI actors (Weidinger et al 2023). AI labs and AI regulators, while also responsible to a degree, allegedly lack the specialist knowledge gained from being embedded in particular sectoral contexts. Such an approach may be appropriate for sectors characterized by well-resourced actors clearly incentivized to embrace these context-specific AI safety challenges. For these actors, this approach furthermore promotes a balance between self-regulation, innovation, and meaningful safety measures. However, relying on sectoral AI actors to lead operationalization efforts can also be a problematic arrangement, especially when those sectors are poorly resourced, lack the trained personnel to do such work, or are also "critical" or "high-risk" sectors (EU AI Act 2024). Sectors such as these may be at risk of being "captured" by other, more powerful actors as well.

In this paper, we address a more fundamental problem: the taxonomies of harm intended to support sectoral operationalization of key ethics, safety, and governance principles are too general to be readily operationalized—they also must account for context. Yet, there is no agreement about how that "contextualization" should proceed. In the absence of such guidance, AI actors have elected (consciously or unconsciously) for overly "doctrinary" approaches: "translating" or "transmitting" general taxonomies into local contexts where those frameworks are to be "received" (Galligan 2006). Applying recent human rights scholarship to this problem, we contend that "reception" is epistemologically insufficient for an effective diffusion of AI safety principles and artifacts, such as taxonomies of harm. Translating them into local contexts as doctrine, just as with human rights, over-general taxonomies of harm and "present[s] them as …single, universal, and immutable concept[s, which] ignores their complexity and by extension only serves to weaken them" (Alston 2024, Merry 2006).

In order for taxonomies of harm to be most useful, and most responsive to the local contexts in which harms manifest, we maintain they must be "vernacularized" (Merry 2006), a process that differs from "translation" in ways we discuss below. We utilize concepts from social theory and the anthropology of law to critically reassess the assumption found in many existing approaches to AI sociotechnical safety that social space is uniform. The degree to which the values and norms of such spaces can differ from those operating in society in general has been underappreciated. Instead, we substitute the notion of a dynamic social ecology, one in which clusters of values, norms, and cultures can be concentrated, stretched, and "refracted," (Abidin 2021) each eddy and whorl forming unique spaces within societies.

With this vision in mind, the problem of operationalization becomes much more apparent: if context constructs harms, then an awareness of the unique norms, cultures, and values operating in a particular sector is essential to understanding which harms are constructed, and how, precisely, they may manifest. From this foundation, we argue that, in order to facilitate the operationalization of novel, robust ethics and safety review methodologies, such as holistic evaluations, the taxonomies of harm underpinning them must also be vernacularized into the context of the sector in which they are to be deployed. In this sense, we elaborate on Mark Goodale's description of a practice of "anthropological ethics" that vernacularization relies upon in order to achieve meaningful success (Goodale 2024). To illustrate this point further, we briefly examine one such social eddy–the space of education, which, incidentally, is precisely the kind of historically-underresourced, "high risk" sector that most urgently needs better solutions to AI safety problems.

Our paper proceeds as follows: First, we discuss why being attentive to the unique norms, values, and discourses that constitute social spaces is essential to operationalizing AI safety principles. We then demonstrate the utility of an anthropological review of an example target space (education), which yields a selection of important values, norms, and discourses that constitute that space. After a brief discussion of the form and purpose of taxonomies of harm in general, we then assess the commensurability of leading general taxonomies of harm that have emerged recently with the education space's unique ethical and social context as established prior. Subsequently, we discuss an example of how valuable a vernacularized taxonomy of harm produced by AI systems in education can be to sector-specific AI ethics. And finally we comment briefly on the general utility of vernacularization in supporting other critical AI safety operationalization efforts across sectors: surfacing hidden stakeholders, enabling local AI actors to carry out meaningful ethical foresight and risk analysis, and promoting a dynamic and community-centered culture of AI safety.

## Understanding Social Spaces: An Anthropological Ethics Approach

### How Are Social Spaces Unique?

The spaces, domains, and sectors for which AI systems are developed are discrete spaces constituted by certain norms and values and an established discourse that continually makes and remakes them. As danah boyd has argued of such "youth spaces" (boyd 2010), we might think of them as "public spheres" (Habermas 1991) in themselves, for, even though the labels of such groupings may have arisen from market logics, they are formed primarily by the "production and circulation of discourses" (Fraser 1990) that are at once unique to them, and also distortions of wider discourses operating in society in general.

Another way we might think about such discrete social spaces is to regard them as socially "exceptional" (Agamben 2005), a construction that depends as much on the operation of peculiar internal logics as it does on the temporally-finite bounds of such logics. Our example, the space of education, is well described as an "exceptional" in this vein: students (and indeed instructors and administrators) go to school and thus enter a space where other of society's norms and values may be suspended or refracted; then, they leave school and resume those temporarily excepted norms and values. Furthermore, such exceptionality is longitudinal: students are students for only a period of time–until graduation, say. As we will see, both of these forms of exceptionality are quite meaningful to AI actors responsible for AI safety in this space.

Understanding that all target spaces vary according to these kinds of dimensions is essential to any effort to vernacularize AI safety principles and artifacts, and particularly taxonomies intended to map how AI systems may cause harm. This is liberating knowledge for all AI actors. Once this principle is understood, the task becomes one of value discovery in keeping with participatory design or constructive technology assessment methods (Rip 2008, Misa et al. 2004). Such knowledge relieves AI system developers of any pretense that they must take up the onerous task of designing principles that purport to be universal. Ultimately, it is an exercise not just in building anthropological knowledge, but also in anthropological ethics, one that forces all AI actors to reconsider where AI safety values are properly produced and described, not unlike the process through which human rights have been normalized via local demands (Goodale 2024).

## An Example: Education's Unique Vernacular

We theorize the education space as one such "refracted public" space (Abidin 2021), and one subject to particularly overt temporal exceptionalities, such as those mentioned in passing above. At first glance, we may perceive striking similarities between the discourses, norms, and values that constitute the education space and those that constitute society in general. But a closer look is needed, for there are many norms and values operating in direct opposition to broader social norms and values. For instance, in the United States, First Amendment rights (which provide for the freedom of speech) do not operate with as wide a remit as they might in American public life more generally. Likewise, students, especially as minors or in public educational institutions, often enjoy additional rights and protections that they may not enjoy once they leave the educational space. Hence an anthropological account of the education space might yield the following results:

**Education Constructs Privacy Differently Than Society in General**

Education spaces are defined by rather different discourses of privacy than operate in wider society. Safe spaces that allow for learners to make mistakes and consequent respect for "public privacy" are valued highly (Nissenbaum 2004; Marwick 2014). Surveillance by legitimate authorities in accord with constitutive ideologies (see below) is tolerated to a larger extent than in wider society and so too is the sharing of sensitive information with certain indirect stakeholders (parents, pastoral staff, administrators) (Hendry and Friedman 2019, Friedman and Nissenbaum 1997).

**Education is Governed by an Ideology of Pedagogy**

Education as a public space is held together by the higher process of learning, and more precisely by the systematic organization of learning by authorities and peers: in other words by the ideology of pedagogy (Pinar 1978). Authorities (teachers, instructors, tutors, administrators) regulate the intellectual, emotional, and behavioral space of education. Given the aim of pedagogy (see below, on transformation), certain rights (e.g. to private property–a mobile phone, a car) can be curtailed legitimately within the education space. Authorities must, however, be perceived as legitimate, most often materialized in their practices of commutative justice (giving grades that suitably correspond to student performance), and in maintaining the relational bonds between them and students, individually and as a whole-class entity.

**Education is Necessarily Sociable**

To many pedagogists, the educational space is necessarily sociable. According to Albert Bandura, the first to articulate Social Learning Theory, learning happens in the interaction of mental processes that are both intrinsic (cognition) and extrinsic (observing the actions and emotional states of others in the same learning environment) (Bandura 1972, 1977). That is, learners take stock of others' reactions to certain educational stimuli before formalizing an understanding themselves (Manz and Sims 1981). Crucially, this social environment may include technology as approaches from Social Computing or Computer-supported Cooperative Work (CSCW) have established (boyd 2010, DiMocco 2008, Leidner 2006).

Furthermore, the relationship between instructor and learner is contingent and not interchangeable and thus has been fruitfully examined from the point of view of care ethics (Freire 1970). Relationships between learners and instructors can be very deep. So too can the relationships between individual learners (Marx and Ko 2012). And recent research suggests the possibility of similar kinds of relationship (even if inappropriate) may come to characterize student-technology interactions (Gabriel et al 2024), a topic about which we will write elsewhere. The unique strengths and persistence of learning relationships directly underpins norms of public privacy and significantly shapes perceptions of legitimate authority.

**Education's Goal Is Social Transformation, Not Social Reproduction**

Many prominent pedagogists maintain that the space of education is meant first and foremost to train individuals to be capable of both self-transformation (Piaget 1976; Vygotsky 1978; Papert 1990) and social transformation (Giroux and McLaren, 2011), not merely the reproduction of society or the production of value. Education's unique power relations and flow (ideologies of pedagogy), sociostructural formations (classrooms, tutorials), and ethical values (public privacy) are each justified by this ultimate goal.

# Taxonomies of Harm Must be Vernacularized to be Operationalized

Having formalized some of the key aspects of the education space's ethical and social vernacular, we turn to an assessment of the utility of taxonomies of harm to support AI safety goals in this context. First we discuss the concept of vernacularization in more detail to demonstrate its applicability to AI ethics and safety operationalization efforts. We then review existing, general taxonomies. Then we discuss a brief example from a vernacularized taxonomy.

**On Existing General Taxonomies of Harm**

In systems design, taxonomies are artifacts used to support efforts to identify, organize, forecast, and reduce harms systems may cause (Banko 2020). Many approaches to responsible system design rely on formal or semi-formal taxonomies to help classify and therefore measure and track progress in mitigating several types of harms. In AI ethics and safety, there are now several existing taxonomies of harms caused by AI systems or frameworks for manual or automated harm assessments built atop them (Shelby 2023, Weidinger et al 2021, Weidinger et al 2022, Magooda et al 2023). Likewise, risk-management approaches to public policy and law utilize taxonomies to implement regulatory frameworks for developing or disruptive technologies (EU AI Act, 2024; National Institute of Standards and Technology 2023).

Although taxonomies aid in reaching safety solutions, they are not AI safety solutions in themselves. Rather, they are essential documentation for delivering certain aspects of an effective AI safety program and can guide both technical and sociotechnical safety assessments. Likewise, they can be used at any stage of system development, from requirements gathering, to development, to post-deployment monitoring. They belong in the toolkit of any responsible AI actor and are key to driving our measurement sciences of AI capabilities and harms.

Although each taxonomy varies according to the priorities of the research team proposing them, the kind of systems for which they are articulated, and the jurisdictional environment in which the system will be deployed, taxonomies of AI-specific harms roughly cohere around the following types of risks:

      I. Discrimination, Hate speech and Exclusion,
      II. Information Hazards,
      III. Misinformation Harms,
      IV. Malicious Uses,
      V. Human-Computer Interaction Harms,
      VI. Environmental and Socioeconomic harms.
      VII. Harms to human systems and institutions

Leading taxonomies typically contain several subclasses and the best are annotated with examples of each harm.

General classifications are essential starting points. They rightly seek to distinguish between types of harms that AI systems can cause, providing needed visibility across the whole "safety surface," so to speak. Each class furthermore suggests technical and sociotechnical mitigation strategies as well (Barocas 2023): for instance, if systems may cause harms by occasionally including hate speech in their outputs, safety strategies should include testing that quantifies the rate at which such undesirable outputs occur, qualitative assessment about the likely impact of those harmful outputs, and a monitoring program that tracks progress towards an elimination of such outputs. But, however well-thought out a general taxonomy of harms caused by AI systems may be, it should begin to be clear even by this brief example that further refinement is needed to account for the specific contours of the intended deployment context.

This is why we argue that it can be useful to think of general taxonomies used in AI safety as analogous to other normative frameworks, such as human rights, in that they are "created through diverse social movements….crystallized…then appropriated by myriad … organizations" that attempt to cultivate them in "terms that make sense in their local communities" (Merry 2017). It is the final step, a process legal anthropologist Sally Engle Merry referred to as

"vernacularization," that is the most critical, for it is only in this step that those frameworks come to meaningfully inhabit the target community "on the ground" (Merry 2006). We elaborate on this point in the next section.

## Vernacularization Resolves the Global-Local Paradox

Vernacularization is something more than translation (Levitt and Merry 2009). Translation seeks to map certain renderings of concepts from one symbolic system to another and is driven by specialists in those symbolic systems (Tymoczko 2014). Those systems, of course, have deep connections to cultures in which they developed and so are never "purely descriptive" (Pym 2008). Instead, the ultimate goal is to arrive at "functional" mappings that are first and foremost stable enough to accomplish the goal of persistent communication and self-representation (Toury 2012). According to metaphor theory and legal hermeneutics, translation also necessarily changes the meaning of each linguistic components being related, merely by placing them in the artificial suspension of mapping (Lakoff, 1992).

On the other hand, vernacularization is necessarily driven by communities themselves. As a term arising from ethnographic and anthropological approaches to human rights law, it has been applied to describe the process a community undergoes as it appropriates appealing external principles (such as the UDHR) and redeploys them in the specific cultural, social, and ethical context of community life. Indeed, some have come to describe vernacularization as a practice—and communities engaging in such practices not as "recipients" shaping doctrinal "transfers" (Bianchini 2021), but as "users" (Janmyr 2022) availing themselves of useful principles in the configuration of their own norms and values. As "users," such communities are also therefore participants in vernacularization (Eslava 2015), hence the characterization of vernacularization as an emerging decolonial approach to cultivating global norms while allowing plural values and discourses to thrive.

While vernacularization, as discussed here, has not yet been taken up in systems design, participatory design methodologies have moved AI ethics and safety, responsible innovation, and governance paradigms towards that practice. A notable example of particular importance to the education space is Google Deepmind's recent LearnLM—a system intended to support learning tasks developed using highly participatory methodologies and treated as a testbed for applying holistic evaluation methods (Jurenka et al, 2024). However, it appears that many developers of AI systems are still operating in the mode of "translation" rather than "vernacularization." "Translation" approaches often encounter a paradox: how can frameworks, which must be standardized as doctrines to serve as effective global AI safety and governance instruments, account for the plural nature of human values across cultures and target deployment spaces? Previous solutions have recognized the inevitability of "tension" (Sorensen 2023) or novel approaches to cultivate "moral imagination" sufficient to overcome a sensation of paradox (Lange 2023).

Adopting a vernacularization paradigm recharacterizes these apparent contradictions as essential features. Novel anthropological approaches to human rights encourages us not to fear a technical mismatch between normative and vernacularized human rights frameworks. This is not because the paradox is simply the "cost of doing business" but rather because the paradox only exists if we insist on conceiving such frameworks as doctrine. Venacularization makes no such assertions, and instead ensures that the principles underpinning normative frameworks come to "reside" meaningfully (Ewick and Silbey 1998) in the practices of a local community, not just its statute book.

As with human rights, so too with AI safety principles recorded in general taxonomies of harm. By relaxing the emphasis on doctrine in favor of an anthropological approach, global AI safety frameworks can become meaningful parts of the responsible AI cultures of appropriating communities (Madhok 2021). In this way does vernacularization actually end up contributing directly to the construction of global AI safety and governance norms more broadly as the sources of a kind of positive law "partly common to all mankind" (Waldron 2012). Most importantly, as a community-driven process, it is the work of vernacularization that stakeholders seeking to effect AI safety and governance solutions even in underresourced spaces such as education can accomplish.

## Towards Vernacularized Taxonomies of Harm

Vernacularization has the clear potential to inform AI ethics and safety efforts, specifically those involving refining general taxonomies of harm for use in unique target contexts. Consider the example below, which shows one plausible result of vernacularizing "Misinformation Harms," which has been defined as "Generating or spreading false, low-quality, misleading, or inaccurate information that causes people to develop false or inaccurate perceptions and beliefs," (Weidinger et al 2023) clearly a critical risk area for education. It may appear that such a definition of risk actually needs no vernacularization. Misinformation, as defined, may in itself be harmful regardless of context. Yet, applying the four values and norms unique to the education space stated above (public privacy, pedagogy, sociability, transformativity), new risk areas, and a new estimation of the magnitude of this risk, vis a vis others, can be reached:

| General Risk Area | Misinformation Harms |
|---|---|
| Definition of Risk | Generating or spreading false, low-quality, misleading, or inaccurate information that causes people to develop false or inaccurate perceptions and beliefs |
| Vernacularized Risk Area | Accuracy, Authenticity and Treatment Harms |
| Education-Specific Risk | Learners may be presented with information that appears factually correct, but either is verifiably incorrect, or is reflective of reasoning that has been rejected. |
| Vernacularized Harms | Basis for cultivating peer-to-peer or peer-to-instructor relationships is weakened<br><br>Trust in formative ideology (pedagogy) is degraded<br><br>Lost opportunity to drive self-transformation<br><br>Reduced capacity to contribute to social transformation |

Table 1: A sample vernacularization of misinformation harms

Misinformation can of course cause serious harm to society in general: it can lead to diminished likelihood that genuine information is believed, or can erode faith in institutions (Shelby et al 2023, Weidinger et al 2023). However, because of the paramount significance of authentic, accurate information to the objective, formative ideology, and socialbility of the education space, harms caused by misinformation are dramatically intensified, and may even be catastrophic.

Applying a holistic model of safety evaluations reveals this in more depth (Weidinger et al 2023). We can see how misinformation harms may appear in different valences *at different levels* of interaction. These longitudinal factors can vary widely across unique spaces, however, and therefore must also be translated into terms meaningful to the education space.

| Vernacularized Harm | Basis for cultivating peer-to-peer or peer-to-instructor relationships is weakened<br><br>Trust in formative ideology (pedagogy) is degraded<br><br>Lost opportunity to drive self-transformation<br><br>Reduced capacity to contribute to social transformation |
|---|---|
| Capability Level | Students may underperform on assessments because they lack factual or process knowledge, or because AI systems are not evaluating student responses appropriately |
| Human interaction Level | Students incur "learning debt" by not having been exposed to relevant, authentic material.<br><br>Underperformance may lead to consequences within (failing a class) and outside (parental discipline, disqualification from clubs or jobs) educational spaces.<br><br>Social bonds between learners and instructors may become strained |
| Social systems | Internal and external trust in learning authorities, technology and tools, or peers may consequently be degraded.<br><br>Students may emerge from formal education poorly prepared to perform desired tasks, leading to disruption of labor market and economy, or weakening of public institutions |

Table 2: How vernacularized misinformation harms may present at different interaction levels

Neither of these examples are comprehensive. The point is that vernacularizing general taxonomies results in far more meaningful assessments of the ethical and safety risk landscape. And, just as with general taxonomies, each of these vernacularized risks suggests specific responses on the part of those responsible for AI safety in educational spaces. Student underperformance is typically already monitored by instructors, but understanding that such underperformance may be caused by AI systems being used in the delivery or evaluation of learning is an essential component of AI safety in the educational space (GPT surprise paper, 2024).

## Overgeneral Taxonomies Can Compound Potential Harms

It is not simply the case that using a general taxonomy is good, and using a vernacularized one is better. Foregoing vernacularization actually presents new risks of harms difficult to articulate by using general taxonomies alone. Moreover, as our previous example demonstrated, those risks may be significantly intensified by values, norms and discourses operating in the target space.

We must be mindful that harms caused by AI systems are often very personally felt. All the more so when harms arise unintentionally, or by "accident" (Amodei et al 2016). However, recent approaches from trust and safety communities in system design urge us to reconsider the cost of accidents. They are not the product of performance failures, but rather of system design failures, and specifically system safety design (Leveson 2004, Raji and Dobbe, 2020). When accidents happen, they have real personal and moral cost (Perrow 1984), and it is these costs that taxonomies of harm try to classify in order to direct mitigation efforts.

In this vein, vernacularized taxonomies of harm can reveal the potential hidden personal and moral costs of system safety failures not immediately apparent when using general taxonomies. Because they account for the unique social and political authorities that constitute the education space, even the cursory tables above enumerate education's likely "moral crumple zones"—a term that originates from accident sciences and is used to describe actors or components that take the blame for any failures in system design (Nissenbaum 1996, Elish 2019). In the case of accuracy, authenticity, and treatment (misinformation) harms caused by AI systems that are materialized as lost opportunities for self-transformation, while students will suffer adverse consequences, it is educators themselves who will likely be held primarily responsible (Sims 2017). Failing to understand the unique ideological centrality of instructor-student relationships not only to the method of pedagogy but also the orientation of education's values and discourse risks missing the very real effect AI safety failures may have on educator efficacy, wellness, and supply.

Accuracy, authenticity, and treatment harms produced by AI systems may also severely degrade educational sociability, as students may also become moral crumple zones if the basis for peer-to-peer support–working with similar, high-quality information–is eroded. Moreover, failure to understand education's unique norm of "public privacy" may also lead to significant clashes between values prioritized by students, instructors, and system designers: learners being surveilled by AI systems that report their time-on-task or areas of suggested improvement may feel such surveillance intrudes on the trusting relationship formed between students or between student and instructor (Crooks 2019, Nissenbaum 2004, Marwick 2014, Henry 2022) eroding again a key value of educational spaces.

We should also note that this is no longer a theoretical question (Williamson 2024). AI systems are being deployed into educational spaces across the globe with very little (if any) effort to vernacularize AI ethics and safety principles or frameworks such as the taxonomies of harm discussed above. The way in which they are raises new questions that underscore the urgency with which this problem. While many educators and administrators are proceeding cautiously with these implementations, others have moved quickly, creating the conditions for serious risk that exceeds the bounds of their isolated experimentation. As a result, we are beginning to develop empirical understandings of the efficacy of AI systems deployed in educational contexts, and early indications suggest that while AI systems may promote higher rates of educational attainment, that may come at the cost of student engagement (Nie et al 2024).

One example stands out: the deployment of AI systems to "remedial" learners, often treated as exceptional zones within already exceptional educational spaces (Watters 2023). There are emerging studies that suggest that exceptional circumstances may be benefited by exceptional solutions, such as AI-based systems that are hyper-personalized (Henkel 2024, French Ministry of Education 2024). Then again, critics of experimental personalization of education serviced through AI-proctored learning experiences maintain that such systems have sometimes served to widen divides between "remedial" and "traditional" learners, putting them on different tracks that increasingly diverge as students progress through their formal education (Holmes 2018).

In any case, we contend that AI actors in the education space deploying AI systems without undergoing vernacularization of general taxonomies likely do not have a clear view of the AI safety surface. High-profile "accidents" will not only incur real personal costs to those likely to be held accountable for AI system failures (justly or unjustly), they may also lead to undetected systemic harms that only manifest in years to come. Ultimately, such accidents may also lead to over-regulation of AI systems responsibly developed and responsibly governed by AI actors in education, preventing the potential benefits AI technologies may offer from being fully realized in the education space.

## Vernacularization as a General AI Safety Operationalization Methodology

We have proposed vernacularization not simply as a deeper analysis of taxonomies of known harms AI systems can cause. It is an inherently participatory AI safety methodology, one that accommodates both plural and shared (but not necessarily universal) values. It allows AI actors inhabiting target spaces to retain their agency in supporting such a multidisciplinary process as AI safety must be. In this respect, it is also grounded in decolonial approaches (Mohamed et al 2020, Muldoon 2023). But vernacularization need not only concern itself with appropriating and recapitulating taxonomies and frameworks needed to operationalize AI safety. Indeed, as one mode of an anthropological ethics, it can also support other critical aspects of context-specific risk evaluation methodologies. We mention some of those in passing:

### Ethical and Risk Forecasting

Vernacularization is itself a form of ethical foresight and risk assessment (Floridi 2021), in that its participants identify areas where existing taxonomies fail to account for known harms AI systems may cause in context. Participants also identify novel risks not accounted for in local discourses, as these may be revealed only in context.

In the education space, for example, vernacularized harms revealed that AI systems deployed to education may lead to novel conceptualizations of "social transformation" itself. No doubt this process entails the transformation of education and a reconfiguration of critical discourses and norms—a process that will involve conflict.

### Stakeholder Engagement

Vernacularization can help identify stakeholders and reveal how direct and indirect stakeholders within a target sector are related to one another (Hendry and Friedman 2019). Discovery on this front can yield invaluable information about the structural features of a target context. This in turn shapes how the vernacular can affect the global or normative.

Again, considering the example of education, to what extent have indirect stakeholders (passers-by, parents, friends) been considered whose data, likenesses, or behavior is collected by AI systems intended for educational spaces?

### Local Value Discovery and Value Balancing

Values and norms are in constant states of becoming, and as such are never fully stable at any given time. Vernacularization, whether of taxonomies or of other practices or principles can prompt communities to reflect critically on their own values, which may lead them to discover new ones or new needs to balance existing ones (Bergman et al 2024).

In an educational space, an admittedly frantic search for strategies to deal with student academic dishonesty (cheating or plagiarism) has led to a critical reassessment of the institution of the assessment. A discourse of "going medieval"—requiring oral presentations or in-class written submissions has grown up in response.

### Scalability and Generalizability

Vernacularization as a methodology can scale to encompass a variety of responsible AI practices and cultures, within sectors and even across sectors.

This makes it especially attractive to underresourced sectors that are not highly centralized, such as education, where AI safety practices and policies can differ from state to state, city to city, or even school to school.

## Limitations

Vernacularization may not be suitable, or feasible, in all cases. It requires greater effort amongst research teams supporting AI system developers. That effort can be costly, both in terms of time and financial resources. However, failing to property resource technology development in general, and AI systems in particular, carries with it its own risks that have been discussed above. Not least of these is the potential to cause substantial harms.

Yet there are still limitations. One critical limitation may be in the research capacity of AI system developers themselves. As a participatory process that requires AI system developers to act as participant observers, vernacularization is dependent on access to key stakeholder communities. Many of these communities, especially those that have historically been minoritized or targeted (such as undocumented immigrants, people of color, or LGBTQIA+ people), may be reluctant to grant such access due to past collective trauma experienced in other research processes (see, for example, Schaff 2010).

It is therefore critical for research teams at AI labs intending to develop or deploy AI systems to any context to include research staff trained in such methods. This implies a financial obligation of these AI actors to support this research process—something that may appear to conflict with business objectives (though, we argue elsewhere, actually makes for more trustworthy and more useful technologies in general). There are certainly other limitations, and we look forward to exploring these in future work.

## Conclusion

In this paper, we have argued that, just as "principles alone cannot guarantee ethical AI," (Middelstadt 2019) general taxonomies alone, however robust, cannot necessarily serve as the basis for operationalizing those principles in particular sectors *by themselves*. In fact, AI actors relying on general taxonomies of harm to operationalize AI ethics and safety principles or strategies, including evaluations and monitoring, may indeed create additional harms. Vernacularization of such general taxonomies is an essential exercise, we maintain. They do not replace general taxonomies of harm (which, of course, have many other critical uses in framing law or policy). However, vernacularized taxonomies have unique value because they respond to the specific normative attributes of target contexts, which we urge systems designers and other AI actors responsible for AI safety to treat as a unique ethical and social spaces in their own right. We also articulated some ways in which vernacularization, as an exercise in anthropological ethics, can also support other AI ethics and safety operationalization efforts– namely by functioning as a kind of ethical foresight, stakeholder discovery, and value balancing exercise. To demonstrate each of these points, we referred regularly to the example of the education space, though, we argue that vernacularization is a generalizable process that AI actors in every sector should include in their AI safety operationalization efforts.